# On the gravitational field of a point-like body immersed in quantum vacuum

Dragan Slavkov Hajdukovic

INFI, Cetinje, Montenegro

dragan.hajdukovic@cern.ch

**ABSTRACT**

Quantum vacuum and matter immersed in it interact through electromagnetic, strong and weak interactions. However, we have zero knowledge of the *gravitational properties* of the quantum vacuum. As an illustration of possible fundamental gravitational impact of the quantum vacuum, we study the gravitational field of an immersed point-like body. It is done under the *working hypothesis*, that *quantum vacuum fluctuations are virtual gravitational dipoles* (i.e. two gravitational charges of the same magnitude but opposite sign); by the way, this hypothesis makes quantum vacuum free of the cosmological constant problem. The major result is that a point-like body creates a halo of the polarized quantum vacuum around itself, which acts as an *additional* source of gravity. There is a maximal magnitude $g_{qvmax}$, of the gravitational acceleration that can be caused by the polarized quantum vacuum; the small size of this magnitude ($g_{qvmax} < 6 \times 10^{-11}\,m/s^2$) is the reason why in some cases (for instance within the Solar System) the quantum vacuum can be neglected. Advanced experiments at CERN and forthcoming astronomical observations will reveal if this is true or not, but we point to already existing empirical evidence that seemingly supports this fascinating possibility.

## 1  Introduction

So far, we had two scientific revolutions in our understanding of gravitation: Newton's law and Einstein's General Relativity. Whatever happens in the future, these two revolutions will remain among the greatest achievements of theoretical physics and the human mind. However, both theories have a wrong assumption in common. The wrong assumption is that matter of the Universe exists in classical, non-quantum vacuum; not surprising because both theories were developed before the existence of quantum vacuum was established.

The quantum vacuum is an essential part of the Standard Model of Particles and Fields (Aitchison 2009). If you are not familiar with the quantum vacuum, just consider it as a new state of matter-energy, radically different from more familiar states (solid, fluid, gas, ordinary plasma, quark-gluon plasma…) but as real as they are. You may imagine quantum vacuum as an omnipresent fluid composed of quantum vacuum fluctuations, or, in a more popular wording, composed of virtual particle-antiparticle pairs with an extremely short lifetime (for instance the lifetime of a virtual electron-positron pair is of the order of $10^{-21}$ seconds).

In the study of electromagnetic, strong and weak interactions, the quantum vacuum *cannot* be neglected; there are well-established *non-gravitational interactions between the quantum vacuum and matter immersed in it*. Let us mention just 3 fascinating, illuminating and experimentally confirmed phenomena from Quantum Electrodynamics.

*First*, the quantum vacuum has a *permanent* tiny impact (but impact!) on "orbits" (i.e. energy levels) of electrons in atoms (Aitchison 2009); this phenomenon is known under the name "Lamb shift".

*Second*, under the influence of a sufficiently strong, external electromagnetic field, quantum vacuum fluctuations can become polarized. A simple "mental picture" of this phenomenon is as follows. A virtual electron-positron pair (i.e. two electric charges of the opposite sign) is in fact a virtual electric dipole, and, from the point of view of Electrodynamics, quantum vacuum is an "ocean" of randomly oriented electric dipoles; a strong electric field can force (can impose) the alignment of these dipoles. Hence, the electric polarization of the quantum vacuum is as real as the analogous polarization of a dielectric. Charged particles (electrons, positrons, protons …) create a microscopic halo of the polarized quantum vacuum around themselves; the effect of this halo is the "screening" of the electric charge of particle. Consequently, if you measure the electric charge of an electron outside of its halo of the polarized quantum vacuum you will get the familiar constant value ($1.602 \times 10^{-19} C$); however if you measure inside the halo (where screening is smaller) you will measure a greater electric charge (L3 Collaboration 2000). At this point it can be useful to look at Figures 2 and 3 and reinterpret them from the point of view of electric dipoles.

*Third*, quantum vacuum fluctuations can be converted into real particles; we can create something from apparently nothing. In fact, 8 years ago (Wilson et al. 2011), the dynamical Casimir effect (i.e. creation of photons from the quantum vacuum) was confirmed; poetically speaking, for the first time we have created light from darkness. By the way, during the next decade, hopefully, we can expect a confirmation of the Schwinger mechanism (Schwinger 1951), i.e. creation of electrons and positrons from the quantum vacuum.

The open question is, *if there are also gravitational interactions between the quantum vacuum and the immersed matter*? Whatever the answer is, the lesson learned from the cosmological constant problem (Weinberg 1989) is that we are missing something very fundamental in our understanding of gravity. The essence of the cosmological constant problem is that, according to our current understanding of gravity, the quantum vacuum, established in the Standard Model of Particles and Fields, must produce gravitational effects many orders of magnitude larger than is permitted by the empirical evidence. Just as a frapping illustration, if we take mass of a neutral pion as a typical mass of quantum vacuum fluctuations in Quantum Chromodynamics, the quantum vacuum within the Earth's orbit around the Sun should act as about $10^{18}$ Solar masses. Quantum vacuum behaves as if its mass-energy is many orders of magnitude larger than its gravitational charge. Despite many scientific papers devoted to the cosmological constant problem, we still do not know why quantum vacuum, apparently, does not respect our prescribed truth that mass-energy and gravitational charge must be the same quantity.

In this brief paper, we consider the simplest (but fundamental) case of a point-like body that is immersed in the quantum vacuum, under a simple (but striking) working hypothesis that, by their nature, *quantum vacuum fluctuations are virtual gravitational dipoles* (i.e. each fluctuation is composed of two gravitational charges of the same magnitude but opposite sign). By the way, one positive and one negative gravitational charge within a fluctuation cancel each other; consequently, the total gravitational charge of the quantum vacuum is zero and (after many sophisticated efforts that have failed) this might be a trivial solution to the cosmological constant problem.

Within the next decade, the hypothesis of virtual gravitational dipoles will be confirmed or rejected by empirical evidence. However, even assuming that hypothesis is wrong, it is still useful to have an example of the impact caused with the replacement of the featureless vacuum by a physical vacuum which acts as a source of gravity. It is important to have, at least one illustrative example, to what extent Newton's law (Eq. (1)) is incomplete if the physical vacuum is neglected.

## 2  Point-like source of gravity – Newton and General Relativity

The first revolution in our understanding of gravity was the Newtonian law. According to Newton, the gravitational field around a non-rotating spherically symmetric body of mass $M_b$ can be described by the gravitational acceleration $\mathbf{g}_N$ (which is in fact the strength of the gravitational field):

$$\mathbf{g}_N = -\frac{GM_b}{r^2}\mathbf{r}_0, \qquad (1)$$

where $\mathbf{r}_0$ and $G$ denote respectively the unit vector and the universal gravitational constant. The story about this discovery gives credit to Newton and an apple that fell on his head, but it is a big injustice that bees, without which apples would not exist, are excluded from the story.

In General Relativistic Gravity (in this particular case we can call it Schwarzschild's gravitation) everything is described with the Schwarzschild metric:

$$ds^2 = c^2\left(1 - \frac{R_S}{r}\right)dt^2 - \left(1 - \frac{R_S}{r}\right)^{-1}dr^2 - r^2(d\theta^2 + sin^2\theta d\varphi^2); \quad R_S \equiv \frac{2GM_b}{c^2}. \qquad (2)$$

While equations (1) and (2) look quite different and they have fundamentally different physical principles behind them, Newton's gravitation is the limit of Schwarzschild's gravitation for large distances from mass $M_b$ (the so-called region of weak field).

Newton's gravitation is invalid theory in the case of a strong gravitational field, for instance it cannot describe black holes. However, also in a relatively weak field there are some tiny, but observable differences showing that General relativity is a better approximation than Newton's theory; for instance you may be familiar with the fact that Newton's gravitation and Schwarzschild gravitation predict slightly different perihelion precession of the planets, with the biggest difference in the case of Mercury because it moves in a stronger gravitational field than the other planets.

Equations (1) and (2) are valid for a body immersed in a gravitationally *featureless* classical vacuum. Consequently, two different observers at distances $r_1$ and $r_2$, measuring the total gravitational charge inside the corresponding spheres will measure the same value $M_b$ as an observer on the surface of the body. As we will show below, everything is radically different if quantum vacuum contribution is included. We do not question the validity of equations (1) and (2), we do not attempt to modify Newton's or Schwarzschild's gravitation; we simply add the quantum vacuum "enriched" with virtual gravitational dipoles as the heretofore neglected source of gravity.

## 3  What if quantum vacuum fluctuations are virtual gravitational dipoles?

Let us introduce the working hypothesis that, by their nature, *quantum vacuum fluctuations are virtual gravitational dipoles*.

Apparently, the simplest and the most elegant realization of this hypothesis is, if particles and antiparticles have the gravitational charge of the opposite sign; of course, nature may surprise us with a different realization of the gravitational dipole-like behaviour of the quantum vacuum.

Let us underscore that so far, there is no empirical evidence that can be cited to disprove the above working hypothesis; on the other hand, there are theoretical arguments against the existence of negative gravitational charge and against the existence of virtual gravitational dipoles (See Appendix A, which presents the main theoretical argument against virtual gravitational dipoles). Apart from Appendix A, the philosophy of the current paper is that it is more important and productive to reveal the physical consequences of the hypothesis than to enter into the exchange of purely theoretical arguments against and for the existence of virtual gravitational dipoles. In a way, the plausibility of consequences can also be considered as an argument in favour of the working hypothesis, but of course, as always in physics, the last word belongs to experiments.

When antimatter is in question, let us underline the experimental fact that particle and its antiparticle have *the same* inertial mass; hence, we do not say that antiparticles have negative mass but that they might have a *negative gravitational charge*, i.e. we assume that the gravitational charge and the inertial mass of an antiparticle have the opposite sign $\bar{m}_g = -\bar{m}_i$ (bar denotes antiparticle), while for particles remains $m_g = m_i$.

This is the right moment to underscore that there are impressive experimental efforts to reveal the gravitational properties of antimatter. Three competing experiments at CERN [ALPHA (Bertsche 2018), AEGIS (Brusa et al. 2017) and GBAR (Perez et al. 2015)] work on the measurement of the gravitational acceleration of antihydrogen (a system composed of an antiproton and an antielectron) in the gravitational field of the Earth. In addition to these already active experiments, a few different experimental groups plan to test antimatter gravity in the leptonic sector of the Standard Model. Positronium (a system composed from an electron and an antielectron) and muonium (an exotic atom made of an antimuon and an electron) may be appropriate systems to test the first and second generation of leptons, respectively (Cassidy & Hogan 2014, Phillips 2018).

Apparently, the ALPHA-g experiment at CERN would be the first one in human history to reveal if antimatter falls up or down; the experimental answer is expected at the end of 2021. If antihydrogen falls up it would be the unprecedented scientific revolution; if it falls as ordinary matter we will know that WEP (the weak equivalence principle) is valid for both matter and antimatter and that the biggest mysteries of contemporary physics, astrophysics and cosmology are not related to the gravitational properties of antimatter. By the way, ALPHA is an extremely successful if not the best antimatter experiment of all time. In 2010, the ALPHA collaboration achieved the first-ever trapping (ALPHA Collaboration 2011) of cold antihydrogen atoms; a seminal success, opening a new era in the study of antimatter. From that time, for the ALPHA team, production and trapping of antiatoms has become routine, making possible a long-waiting spectroscopy of antihydrogen (ALPHA Collaboration 2017) as a fundamental tool to look for the eventual differences between matter and antimatter.

Let us turn back to theoretical considerations. According to our hypothesis we consider a quantum vacuum fluctuation (See Figure 1) as a system of two gravitational charges of the opposite sign; consequently, the total gravitational charge of a vacuum fluctuation is zero, but it has a non-zero gravitational dipole moment $\boldsymbol{p}_g$

$$\boldsymbol{p}_g = m_g \boldsymbol{d}, \quad |\boldsymbol{p}_g| < \frac{\hbar}{c}, \qquad (3)$$

Here, $m_g$ denotes the magnitude of the gravitational charge, while, by definition, the vector $\boldsymbol{d}$ is directed from the antiparticle to the particle and has a magnitude $d$ equal to the distance between them. The inequality in (3) follows from the fact that the size $d$ of a quantum fluctuation is smaller than the reduced Compton wavelength (i.e. $d < \lambdabar_g = \hbar/m_g c$).

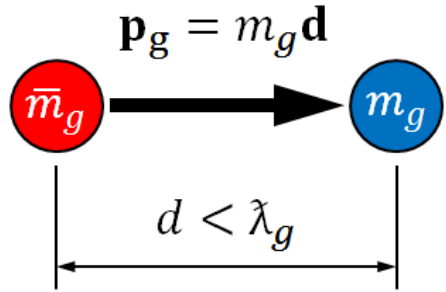

**Figure 1.** A virtual gravitational dipole is defined in analogy with an electric dipole: two gravitational charges of the opposite sign ($m_g > 0, m_g + \bar{m}_g = 0$) at a distance $d$ smaller than the corresponding reduced Compton wavelength $\lambdabar_g$.

If gravitational dipoles exist, the gravitational polarization density $\boldsymbol{P}_g$, i.e. the gravitational dipole moment per unit volume, can be attributed to the quantum vacuum. It is obvious that the magnitude of the gravitational polarization density $\boldsymbol{P}_g$ satisfies the inequality $0 \leq |\boldsymbol{P}_g| \leq P_{gmax}$ where 0 corresponds to the random orientations of dipoles, while the maximal magnitude $P_{gmax}$ corresponds to the case of saturation (when all dipoles are aligned with the external field). The value $P_{gmax}$ must be a universal constant related to the gravitational properties of the quantum vacuum. Later we will discuss the possibility of the experimental determination of the eventual universal constant $P_{gmax}$.

If the external gravitational field is zero, quantum vacuum may be considered as a fluid of randomly oriented gravitational dipoles (Figure 2). In this case everything is equal to zero: the total gravitational charge, the gravitational charge density and the gravitational polarization density $\boldsymbol{P}_g$. Of course, such a vacuum is not a source of gravitation.

Fortunately, the random orientation of virtual dipoles can be broken by the gravitational field of the immersed Standard Model matter. Massive bodies (particles, stars, planets, black holes…) but also many-body systems such as galaxies are surrounded by an invisible halo of the gravitationally polarized quantum vacuum, i.e. a region of non-random orientation of virtual gravitational dipoles (Figure 3).

The magic of non-random orientation of dipoles, i.e., the magic of the gravitational polarization of the quantum vacuum is that the otherwise gravitationally featureless quantum vacuum becomes a source of gravity! Of course, the gravitational polarization of the quantum vacuum has no impact on the real gravitational charge and the gravitational charge density, but, in the region of polarization, the gravitational polarization density $\boldsymbol{P}_g$ is not zero. If you switch off the external gravitational field, you have random orientation of dipoles, i.e. $\boldsymbol{P}_g = 0$. If you switch on the gravitational field, in the region of polarization you have non-random orientation of dipoles, i.e. $\boldsymbol{P}_g \neq 0$.

$\boldsymbol{P}_g = 0$          $\boldsymbol{P}_g \neq 0$

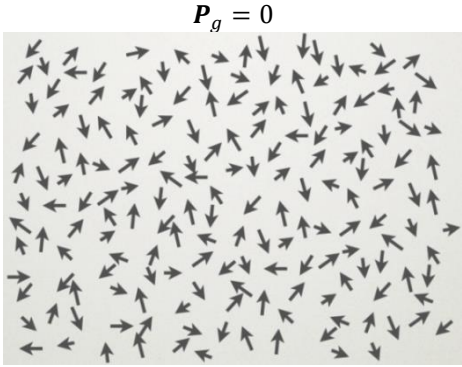 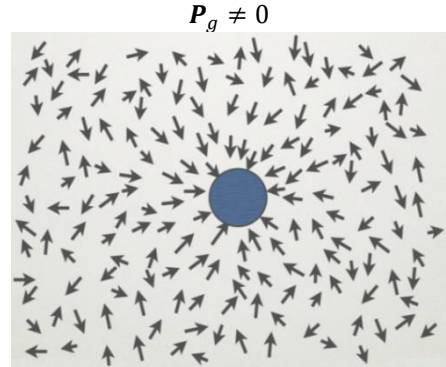

**Figure 2.** Randomly oriented gravitational dipoles (in absence of an external gravitational field).

**Figure 3.** Halo of non-random oriented gravitational dipoles around a body with baryonic mass $M_b$.

The *spatial variation* of the gravitational polarization density generates (Hajdukovic 2011; Hajdukovic 2014) a *gravitational bound charge density* of the quantum vacuum

$$\rho_{qv} = -\nabla \cdot \mathbf{P}_g , \qquad (4)$$

You can consider this gravitational bound charge density as an *effective gravitational charge density*, which acts as if there is a real non-zero gravitational charge. That is how the magic of polarization works; *quantum vacuum is a source of gravity thanks to the immersed Standard Model matter*.

If you are familiar with Maxwell's electrodynamics you will recognize the *full analogy* with the fundamental relation $\rho_e = -\nabla \cdot \mathbf{P}_e$ between the electric bound charge density $\rho_e$ and the electric polarization density $\mathbf{P}_e$.

Only future empirical evidence can tell us if the above relation is correct or wrong.

Let us end this section with an attempt to estimate, from the microscopic point of view, the numerical value of the presumed universal constant $P_{gmax}$.

There are many kinds of quantum vacuum fluctuations. For simplicity imagine that there is only one kind of dipoles (which produce the same effect as all different kinds of dipoles together). The number density of fluctuations is known to be $1/\lambda_g^3$ (where $\lambda_g = h/m_g c$ denotes the Compton wavelength), while the magnitude of individual dipole moments is a fraction of $\hbar/c$. Hence

$$P_{gmax} = \frac{A}{\lambda_g^3}\frac{\hbar}{c} \equiv \frac{A}{2}\frac{m_g}{\pi \lambda_g^2} \qquad (5)$$

where $A < 1$ is a dimensionless constant. We assume that $A = 1/2\pi$; without entering details, this choice (Hajdukovic 2014) assures compatibility of our results and the Unruh temperature derived within the framework of General Relativity. In order to get an idea about the value of $P_g$, let us use mass of a neutral pion $\pi^0$ which is $2.4 \times 10^{-28} kg$. Don't be misled; we do not attribute any crucial importance to $\pi^0$. What is most important is that $\pi^0$ represents a typical mass in Quantum Chromodynamics and since the time of Dirac several intriguing coincidences (Dirac 1937, Dirac 1938, Weinberg 1972, Hajdukovic 2010, 2014) were related to its mass. With this choice, from Eq.(5) we have an interesting estimate: $P_{gmax} \approx 0.072 \, kg/m^2$, or in units preferred par astronomers $P_{gmax} \approx 34 M_{Sun}/pc^2$.

Of course, only experiments can reveal the exact value of $P_{gmax}$. Based on experience with empirical data, our guess is that the true value of $P_{gmax}$ is slightly smaller than this estimate that will be used as a working value for numerical illustrations.

## 4 A point-like body immersed in the quantum vacuum

In the case of spherical symmetry, the fundamental equation (4) that determines the effective gravitational charge density of the quantum vacuum reduces to:

$$\rho_{qv}(M_b, r) = \frac{1}{r^2} \frac{d}{dr}[r^2 P_g(M_b, r)], P_g(M_b, r) \equiv |\boldsymbol{P}_g(M_b, r)| \geq 0, \tag{6}$$

Let us note that from purely mathematical point of view, density $\rho_{qv}(M_b, r)$ can be positive, negative and zero. The effective gravitational charge density is positive in a region in which $r^2 P_g(M_b, r)$ increases with $r$, and negative in a region in which $r^2 P_g(M_b, r)$ decreases with $r$. It can be zero only in a region in which $r^2 P_g(M_b, r)$ has a constant value.

Eq. (6) leads to the following *effective gravitational charge of the gravitationally polarized quantum vacuum* within a sphere of radius $r$:

$$M_{qv}(M_b, r) = \int_0^r \rho_{qv}(M_b, r) \, dV = 4\pi r^2 P_g(M_b, r), \tag{7}$$

The effective gravitational charge $M_{qv}(M_b, r)$ must have an upper bound $M_{qvmax}(M_b)$ to which it tends asymptotically in the limit $r \to \infty$; simply, after a characteristic size $R_{ran}$ the random orientation of dipoles dominates again, because the gravitational field is not sufficiently strong to perturb random orientation. This is both in agreement with our intuition and with the experimental fact about halos caused by the electric polarization of the quantum vacuum (halos are limited both in the size and in the content of the effective electric charge).

Now, according to Newton's law and Eq. (7), the gravitational acceleration caused by the quantum vacuum around a point-like body is determined by

$$\boldsymbol{g}_{qv}(M_b, r) = -4\pi G P_g(M_b, r) \boldsymbol{r}_0, \tag{8}$$

Function $P_g(M_b, r)$ is not known but it has an exact upper limit that, in principle, can be measured. Namely, as already mentioned within Section 3, in the region of saturation (which is roughly a sphere of radius $R_{sat}$ that will be estimated bellow

$$P_g(M_b, r < R_{sat}) \approx P_{gmax} \text{ and } P_g(M_b, r \ll R_{sat}) = P_{gmax}, \tag{9}$$

Hence, as a trivial consequence of equations (6), (7) and (8), sufficiently deep inside the region of saturation, we have robust results which do not depend on the exact form of function $P_g(M_b, r)$, and, additionally, do not depend on the central mass $M_b$.

$$\rho_{qv}(r) = \frac{2P_{gmax}}{r}; \, M_{qv}(r) = 4\pi P_{gmax} r^2; \, \boldsymbol{g}_{qv}(r) = -4\pi G P_{gmax} \boldsymbol{r}_0, \tag{10}$$

The last of the above equations is a fundamental (and in principle testable) prediction of the enormous importance. There is a *maximal magnitude* $g_{qvmax} = 4\pi G P_{gmax}$ of the gravitational acceleration that can be caused by the quantum vacuum; this magnitude is a universal constant. If we use the working value, $P_{gmax} = 0.072 \, kg/m^2$ we have $g_{qvmax} = 6 \times 10^{-11} \, m/s^2$.

A brief digression. Astronomical observations have revealed that in galaxies, the Newtonian acceleration caused by the existing Standard Model matter (i.e. matter made of quarks and leptons interacting through the exchange of gauge bosons), is only a *fraction* of the total observed acceleration (i.e. $g_N/g_{tot} < 1$). This is an empirical fact independent of theoretical attempts to explain it by dark matter (Peebles 2017) or MOND (for a review see Famaey & McGaugh 2012). This phenomenon is significant only when the Newtonian gravitational field is very weak, *roughly speaking only a few times stronger than our value for $g_{qvmax}$*. The open question is if this is just a surprising coincidence or a hint that the quantum vacuum acts as a source of gravity.

The second testable prediction is that $M_{qv}[r]$ is a parabolic function; hence, there is a constant surface density $M_{qv}(r)/4\pi r^2 = P_{gmax}$.

Now, using the maximal magnitude $g_{qvmax}$, we can estimate the size of the region of saturation. It seems reasonable to assume that saturation is a dominant phenomenon only in the region in which the magnitude $g_N$ of the Newtonian acceleration is larger or equal to the maximal magnitude $g_{qvmax}$ that can be caused by the quantum vacuum. Hence, as a working definition of $R_{sat}$ we have:

$$g_N \geq g_{qvmax} \Rightarrow R_{sat} = \sqrt{\frac{M_b}{4\pi P_{gmax}}}, \quad (11)$$

Let us give two numerical examples. For a single proton, Eq. (11) gives $(R_{sat})_p \approx 4.3 \times 10^{-14} m$. For the Sun, $(R_{sat})_{Sun} \approx 1.5 \times 10^{15} m$; roughly $10^4 AU$.

According to Eq. (7), an observer at a distance $r$ from the point-like body, measures the mass of the body plus the effective gravitational charge of the quantum vacuum within the corresponding sphere of radius $r$, i.e. $M_{tot}(M_b, r) = M_b + M_{qv}(M_b, r)$, or more explicitly

$$M_{tot}(M_b, r) = M_b + 4\pi r^2 P_g(M_b, r), \quad (12)$$

The key prediction of Eq. (12) is that two observers at different distances $r_1$ and $r_2$ measure different central masses, i.e. $r_2 > r_1 \Rightarrow M_{tot}(M_b, r_2) > M_{tot}(M_b, r_1)$. In general (See Figure 4) the function $M_{tot}(M_b, r)$ increases from $M_b$ to its horizontal asymptote $M_b + M_{qvmax}(M_b)$. We already know that in a relatively small central part around the body (region of saturation) total gravitational charge is given by $M_b + 4\pi P_{gmax} r^2$; transition from this parabolic growth to asymptotic behaviour is the most enigmatic part.

Finally, Eq. (12), or equivalently Eq. (8), together with the Newton's law (Eq. (1)), lead to the fundamental result for the gravitational field of a point like body immersed in the quantum vacuum.

$$\boldsymbol{g}_{tot}(M_b, r) = -\frac{GM_b}{r^2}\boldsymbol{r}_0 - 4\pi G P_g(M_b, r)\boldsymbol{r}_0, \quad (13)$$

If correct, Eq. (13) is a third revolution in our understanding of gravity. It differs from Newton's law (Eq. (1)) in the second term on the right-hand side that gives the gravitational contribution of the quantum vacuum. It is obvious that this is not a modification of Newton's law; *Newton's law is valid but quantum vacuum acts as an additional (so far forgotten) source of gravity*.

One major point is that in principle a point-like body is no more a point-like source of gravity, because it is inseparable from the halo of the polarized quantum vacuum around it; a halo that can extend to very large distances (for instance the halo of the Sun is larger than the Solar system).

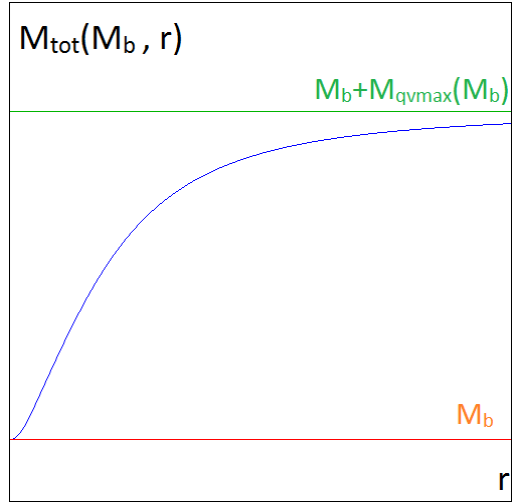

Figure 4: Schematic presentation of total mass measured within a sphere of radius r. Red line is total mass (equal to $M_b$) with the neglected Quantum Vacuum. Blue line is the gravitational charge (mass) with the included Quantum Vacuum that tends asymptotically to $M_b + M_{qvmax}(M_b)$.

Let us underline that a major task is to reveal the exact function $P_g(M_b, r)$. Our current understanding of the quantum vacuum is not enough to rigorously determine this fundamental function; however (and it is already a big step) we can get a rough approximation, which is valid for the whole halo and not only in the region of saturation. As we will see later, an approximation can be obtained from consideration of an ideal system of non-interacting gravitational dipoles in an external gravitational field. Hence, the gravitational polarization of the quantum vacuum is considered as analogous to polarization of a dielectric in external electric field, or a paramagnetic in an external magnetic field! It is very astonishing that it gives an apparently reasonable approximation. We will achieve this in Section 4.2; here we give just the results.

$$P_g(M_b, r) = P_{gmax} \tanh\left(\frac{R_{sat}}{r}\right), \ r < R_{ran}, \tag{14}$$

$$M_{qv}(M_b, r) = 4\pi P_{gmax} r^2 \tanh\left(\frac{R_{sat}}{r}\right), \ r < R_{ran}, \tag{15}$$

We leave discussion of these results for section (4.2). However, let us point out that Eq. (15) in addition to already known parabolic growth in the region of saturation, predicts a linear growth in the outer parts of the halo of the polarized quantum vacuum. More precisely, because for small $x$ $\tanh(x) \approx x$ (for instance already for $x = 1/3$, we have good approximation $\tanh(1/3) = 0.321$), Eq. (15) leads to

$$M_{qv}(M_b, r \gg R_{sat}) = \left(4\pi P_{gmax} R_{sat}\right) r = \left(\sqrt{4\pi P_{gmax} M_b}\right) r, \tag{16}$$

The second equality in (16) is the result of the estimate of $R_{sat}$ given by Eq. (11).
Let us rewrite Eq. (16) using the microscopic interpretation (5) for $P_{gmax}$ and keep in mind that $m_g$ and $\lambda_g$ are close to the mass and the Compton wavelength of neutral pion $\pi^0$.

$$\frac{M_{qv}(M_b, r \gg R_{sat})}{r} = \frac{1}{\sqrt{\pi}} \frac{\sqrt{m_\pi M_b}}{\lambda_\pi}, \tag{17}$$

Hence, we have a striking prediction for the radial gravitational charge density of the quantum vacuum. Find the geometrical mean $\sqrt{m_\pi M_b}$ of mass of a neutral pion $m_\pi$ (i.e. a typical quantum vacuum fluctuation) and the Standard Model mass (usually called baryonic mass) $M_b$ of a body and divide this geometrical mean by the Compton wavelength of fluctuation; what you get is very close to the value of the radial gravitational charge density of the halo of the quantum vacuum.

## 4.1 Regions about a point-like body

Before we continue, let us give one more "mental picture" (See Figure 5) that displays the above results and is complementary to "mental picture" given by Figure 4. You can imagine three regions of quantum vacuum around a body.

The first region (inside a sphere with a characteristic radius $R_{sat}$) is region of saturation. Strictly speaking when $r \to 0$, the function $P_g(M_b, r)$ tends asymptotically to its upper bound $P_{gmax}$, but approximately we can use $P_g(M_b, r) \approx P_{gmax}$ within the whole region. Consequently, $M_{qv}(M_b, r)$ increases as $r^2$ (of course the increase is slightly slower than $r^2$ in the outer part of the region of saturation).

Far from the body (outside of a sphere with a characteristic radius $R_{ran}$) is a region in which the random orientation of dipoles is dominant. Strictly speaking, in the limit $r \to \infty$ the effective gravitational charge density $\rho_{qv}(M_b, r)$ of the polarized quantum vacuum must tend asymptotically to zero. Consequently, according to Eq. (6), the basic function $r^2 P_g(M_b, r)$ tends asymptotically to a constant; more precisely $r^2 P_g(M_b, r) \to M_{qvmax}(M_b)/4\pi$ where we have used notation introduced after Eq. (7).

The key point is that observers from the region of random orientation of dipoles are practically outside of the halo of the polarized quantum vacuum; consequently, with a high accuracy they all measure a central mass $M_{tot}(M_b, r) = M_b + M_{qvmax}(M_b)$ and, in the region of random orientation, they have the correct description of gravity using this mass and Newton's law. In conclusion, the Newtonian law with mass $M_b$ is very accurate deep inside the region of saturation (where the contribution of quantum vacuum can be neglected), and, with mass $M_b + M_{qvmax}(M_b)$ it is again very accurate far away in the region of random orientation (where there is no further increase of the effective gravitational charge of the quantum vacuum).

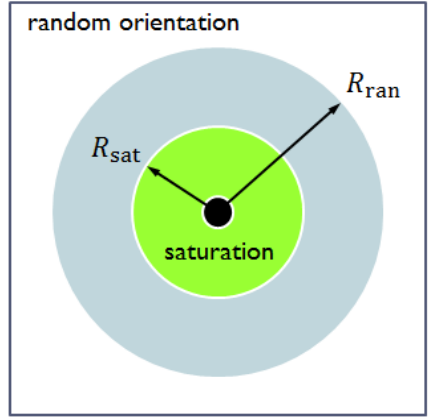

**Figure 5**. Schematic presentation of regions around a body. Region of saturation (green), region of the partial alignment of gravitational dipoles (blue) and region of random orientation (white region after $r > R_{ran}$). The effective gravitational charge $M_{qv}(M_b, r)$ of the polarized quantum vacuum has mainly parabolic growth in the region of saturation, linear growth in the region of partial alignment and asymptotic approach to $M_{qvmax}(M_b)$ in the region of random orientation. The magnitude of the gravitational polarisation density $P_g(M_b, r)$ decreases from the maximal value $P_{gmax}$ (in the region of saturation) towards zero; decrease is respectively as $1/r$ and $1/r^2$ in the region of partial alignment and region of random orientation.

Between the region of saturation and region of random orientation, there is a region (in blue) with a partial (incomplete) alignment of gravitational dipoles. As mentioned after Eq. (6) in this region $r^2 P_g(M_b, r)$ must increase with $r$, and from the point of view of beauty and simplicity it must be a linear function of $r$. If so, the effective gravitational charge density determined by Eq. (6), increases mainly as $r^2$ in the region of saturation, as $r$ in the region of partial alignment and is a constant in the region of random orientation; hence in all regions we have the law of the same form $r^x$ with $x = 2, x = 1\ and\ x = 0$ respectively for regions of saturation, partial polarization and random orientation. This is both beautiful and apparently supported by observations. Namely, at very large distances from the centre, a galaxy can be considered as a point-like body and everything is happening as if the total dynamical mass at a large distance is proportional to that distance.

## 4.2 The simplest approximation for $P_g(M_b, r)$

Let us start with a surprise. Consider a magnetic moment $\boldsymbol{\mu}$ in an external magnetic field $\boldsymbol{B}$, with energy $\varepsilon_\mu = -\boldsymbol{\mu} \cdot \boldsymbol{B} = -\mu_z B$; for simplicity let us limit to the simplest case when magnetic dipoles can have only two energy levels, $\varepsilon_1 = -\mu_z B$ and $\varepsilon_2 = \mu_z B$.

The next step (an easy exercise for students who prepare for an exam in statistical physics) is to find Partition function $Z$ for an ideal system of very large number $N$ of non-interacting dipoles, and after that to find the corresponding magnetization.

For a system composed from non-interacting particles, with each particle having one of two possible energies $-\mu_z B$ or $\mu_z B$, a simple calculation leads to

$$Z = \left[\exp\left(-\frac{\varepsilon_1}{k_B T}\right) + \exp\left(-\frac{\varepsilon_2}{k_B T}\right)\right]^N = \left[\exp\left(\frac{\mu_z B}{k_B T}\right) + \exp\left(-\frac{\mu_z B}{k_B T}\right)\right]^N = \left[2 \cosh\left(\frac{\mu_z B}{k_B T}\right)\right]^N, \quad (18)$$

The average magnetic dipole moment can be easily calculated from the partition function

$$\bar{\mu}_z = \frac{1}{N} k_B T \frac{\partial \ln Z_N}{\partial B} = \mu_z \tanh\left(\frac{\mu_z B}{k_b T}\right), \quad (19)$$

If the number density (i.e. number per unit volume) of dipoles is $n$, magnetisation is

$$M = n\bar{\mu}_z = n\mu_z \tanh\left(\frac{\mu_z B}{k_B T}\right) = M_{max} \tanh\left(\frac{\mu_z B}{k_B T}\right), \quad (20)$$

From a purely mathematical point of view there is no difference between a system of non-interacting magnetic dipoles in an external magnetic field and a system of non-interacting electric dipoles (with electric dipole moment $\boldsymbol{p}_e$ and energy $\varepsilon_e = -\boldsymbol{p}_e \cdot \boldsymbol{E} \equiv -p_{ez} E$) in an external electric field $\boldsymbol{E}$; physical phenomena are different but mathematical equations are the same. In complete analogy with (15) the electric polarization density $P_e$ is:

$$P_e = P_{emax} \tanh\left(\frac{p_{ez} E}{k_B T}\right), \quad (21)$$

The key point is that in this simplest case of non-interacting dipoles both magnetization $M$ and the electric polarisation density $P_e$ are described by a hyperbolic tangent function. If gravitational dipoles exist, together with magnetic and electric dipoles they are in a trio of mathematically identical models; hence the gravitational polarization density $P_g$ must be described by a hyperbolic tangent function!

$$P_g(M_b, r) = P_{gmax} \tanh\left(\frac{p_{gz} g_{tot}(r)}{k_B T}\right), \quad (22)$$

However, it is not clear what is $k_B T$ for the quantum vacuum and consequently what is the ratio of energies $p_{gz} g_{tot}(r)/k_B T$ in the case of gravitation.

It is obvious that equations (14) and (15) follow immediately from Eq. (22) if we impose condition

$$\frac{p_{gz}g_{tot}(r)}{k_B T} \propto \frac{R_{sat}}{r} \, , \tag{23}$$

However, the question remains how, an apparently strange relation as (23) can be possible. Without entering into speculations let us note that relation (23) is possible if we interpret $k_B T$ as energy of gravitational dipoles in an external gravitational field that has fundamental value $g_{qvmax}$, so that $k_B T \approx p_{gz} g_{tot}(R_{sat})$.

An ideal gas of gravitational dipoles is apparently a reasonable approximation practically in the entire halo ($r < R_{ran}$) but it is not surprising that it cannot describe transition to the asymptotic behaviour in the region of random orientation. The major shortcoming of Eq. (15) is that it continues to be a linear function in the region of saturation, instead of tending to a constant. In principle, function (15) must be extended to also cover the region of random orientation. The new function can be written in the form:

$$M_{qv}(M_b, r) = 4\pi P_{gmax} r^2 \tanh\left(\frac{R_{sat}}{r}\right) f(R_{ran}, r) \, , \tag{24}$$

It is obvious that function $f(R_{ran}, r)$ must satisfy 2 conditions. First, in order to preserve Eq. (15) in the domain of its validity, it is necessary to have $f(R_{ran}, r < R_{ran}) \approx 1$. Second, in order to get a constant value in the region of random orientation, $f(R_{ran}, r > R_{ran})$, must be inversely proportional to $r$. A simple interpolating function which satisfies these conditions is

$$f(R_{ran}, r) = \tanh\left(\frac{R_{ran}}{r}\right) \, , \tag{25}$$

Of course, the choice of interpolating function is not unique; the function (25) is just a rough working approximation (or toy model) expected to give correct qualitative behaviour. Let us note that an analogous situation appears in many emerging theories. For instance, there are different interpolating functions in MOND, different empirical laws of distribution of dark matter, and many different functions for the inflation field in the cosmic inflation theory.

Let us note that, for each interpolating function, $R_{ran}$ can be expressed as a function of $R_{sat}$. For interpolating function (25), the asymptotic behaviour $M_{qv}(M_b, r) \to M_{qvmax}(M_b)$ leads to:

$$R_{ran} = \frac{M_{qvmax}(M_b)}{M_b} R_{sat} \, , \tag{26}$$

## 5 Possible Tests in Solar System

At distance of 100AU from the Sun (it is roughly the size of our Solar System), according to Newton law, the gravitational acceleration caused by the Sun is $g_N = 5.9 \times 10^{-7} \, m/s^2$ i.e. at least $10^4$ times larger than the maximal acceleration that can be caused by the quantum vacuum. Hence, if planets and other celestial bodies are neglected, there is a single halo of the polarized quantum vacuum around the Sun and it is the innermost part of the region of saturation. Within a single halo model, the only gravitational effect of the quantum vacuum is a tiny constant acceleration towards the Sun so that the magnitude of the total acceleration towards the Sun is $\boldsymbol{g_{tot} = g_N + g_{qvmax}}$.

However, this simple picture of a single halo is not correct because all planets and smaller celestial bodies have their own halos of the polarised quantum vacuum. For instance, near the Earth, the gravitational field of the Earth is much stronger than the gravitational field of the Sun and other bodies; hence dipoles are oriented towards the Earth, not towards the Sun, and consequently the Earth has its own halo marked by saturation. Of course, corresponding to each halo is an effective gravitational charge; consequently, masses of all celestial bodies are slightly increased by their individual halos. In addition to dipoles that point to a body there are also dipoles aligned with a resultant gravitational field that doesn't point to any Solar System body. Hence, a single halo model is "blind" for two major impacts of the quantum vacuum: *modification of mass of bodies by their individual halos* and *impact of dipoles which do not belong to any individual halo*.

## 5.1 Solar System Ephemerides

It is obvious that solar system ephemerides are an important tool in understanding if the proposed gravitational properties of the quantum vacuum are possible or not. Ephemerides are the result of a numerical integration of the dynamical equations of motion which describe the gravitational physics of the Solar system. Historically, the first ephemerides were created using only Newtonian gravity. Today, ephemerides include General Relativistic effects. What we propose is to include quantum vacuum as well. More precisely, we propose to create new ephemerides, with the quantum vacuum, from the beginning included as a source of gravity, in the dynamical equations of motion. Comparison of "quantum vacuum" ephemerides with the existing ephemerides will reveal if the presumed gravitational impact of the quantum vacuum is compatible or not with the empirical evidence.

In fact, while the motivation was not the quantum vacuum, ephemerides were already used to impose an upper limit on an eventual anomalous constant acceleration towards the Sun. For instance (Fienga 2009) have concluded that a constant acceleration larger than 1/4 of the Pioneer anomaly is incompatible with the observed motion of the planets in Solar System; hence they have established an upper limit of about $2 \times 10^{-10} \, m/s^2$. It is important to underline that this study (and other similar studies (Pitjev & Pitjeva 2013)) correspond to what we have called above "a single halo" model, which neglects the complexity of the impact of the quantum vacuum. Hence, a crucial shortcoming of the mentioned studies is that they used "a single halo" model, together with the existing ephemerides that neglect quantum vacuum in the dynamical equations of motion; the complexity of quantum vacuum effects demands creation of new ephemerides, with the quantum vacuum included in equations of motion from the beginning.

It is obvious that in the Solar System, the fundamental equation $\rho_{qv} = -\boldsymbol{\nabla} \cdot \boldsymbol{P}_g$ has no spherical symmetry. However, creation of "quantum vacuum" ephemerides is facilitated with the fact that within the Solar System the gravitational field is sufficiently strong to produce saturation; hence at each point (if we neglect the insignificant regions in the neighbourhood of the Lagrangian points) we have the same, maximal magnitude of the gravitational polarization density.

## 5.2 A measurement of the universal constant $P_{gmax}$?

Let us imagine an ideal two-body system i.e. an isolated binary composed of two point-like bodies which have sufficiently small mass so that Newton's law is exact description (in the sense that the General Relativistic result and Newton's result differ by a value that is much smaller than the precision of our measurements). While in principle it is not crucial, it can be preferable for the study of the orbit to have a much more massive central body.

The key point is that the orbit in such an ideal system is a fixed ellipse without any kind of precession. It is well known that in the case of General Relativity, i.e. the Schwarzschild metric precession exists even in an ideal binary system; in fact, such an additional precession of the orbit of Mercury was historically the first support for General Relativity. However, as we consider a binary of small mass, general relativistic precession is much smaller than the value that we can observe. Hence, if our theory of gravity is correct, the result of measurement will be precession equal to zero; any non-zero precession would be signature of a new physics.

Precession of an orbit is inevitable if there is a constant acceleration $g_{qvmax}$, caused by quantum vacuum in the region of saturation. More precisely, precession per orbit is:

$$\Delta \omega_{qv} = -2\pi \sqrt{1-e^2} \frac{a^2}{G\mu} g_{qvmax} \equiv -8\pi^2 \sqrt{1-e^2} \frac{a^2}{\mu} P_{gmax}, \qquad (27)$$

Fortunately, precession (27) can be sufficiently large and, if not masked by Newtonian precession, observable for low-mas binaries. Hence, if we know the parameters of the system (eccentricity $e$, semi-major axes $a$ and total mass $\mu$ of the binary) the measurement of precession is equal to the measurement of $P_{gmax}$ and $g_{qvmax}$.

Equation (27) is result of a relatively simple integration of the following equation from the classical celestial mechanics (see for instance the book of Murray & Dermott 1999).

$$\frac{d\omega}{dt} = \frac{\sqrt{1-e^2}}{e}\sqrt{\frac{a}{\mu}}A(r)\cos f \ . \tag{28}$$

In Eq. (28), $f$ denotes the true anomaly, while $A(r) << g_N$ is a tiny perturbation of the Newtonian gravitational field $g_N$; in calculations leading to Eq. (27) we have used $A(r) = g_{qvmax}$.

Of course, ideal binary systems, with Newtonian precession $\Delta\omega_N$ equal to zero, do not exist. In the absence of ideal binaries, the best possibility is to look for a trans-Neptunian binary (Gai & Vecchiato 2014) in which precession caused by the quantum vacuum is bigger than Newtonian precession ($\Delta\omega_{qv} > \Delta\omega_N$).

# 6   Tests in galaxies

Tests in the Solar System are inevitably testing in the region of saturation. In fact, the radius of saturation of our Sun is so big that for larger distances, because of the proximity of other bodies, the gravitational field of the Sun is no longer dominant; in other words the Sun is not sufficiently isolated, not sufficiently far from other bodies and consequently, an external gravitational field prevents the Sun from developing a full halo of the polarized quantum vacuum.

While some point-like bodies can be sufficiently far from other bodies and develop halos much bigger than the region of saturation, systematic appearance of such halos can be expected only around galaxies and larger structures in the Universe. However, when the Universe was smaller (hence structures were closer to each other), nearly full-size halos of the polarized quantum vacuum were not possible. In general, we can expect the growth of halos with the expansion of the Universe.

Of course, a galaxy is not a point-like body, but at very large distances from its centre it can be roughly approximated by a point-like body; consequently, as a preliminary test of the impact of the quantum vacuum, equations (16) and (17) can be compared with the empirical evidence for galaxies at large distances from the centre. For instance, taking $M_{bMW} \approx 6.5 \times 10^{10} M_{Sun}$ for the baryonic mass (Standard Model mass) of our galaxy Milky Way, together with $P_{gmax} \approx 34 M_{Sun}/pc^2$, gives the following value for the constant of proportionality in equations (16) and (17): $\sqrt{4\pi P_{gmax} M_{bMW}} \approx 52.6 \, M_{Sun}/pc$. Consequently, for our galaxy, the effective gravitational charge of the quantum vacuum, and total dynamical mass within radius of 260kpc are respectively $M_{qvMW}(260kpc) \approx 1.37 \times 10^{12} M_{Sun}$ and $M_{totMW}(260kpc) \approx 1.43 \times 10^{12} M_{Sun}$.

This result is in surprising agreement with empirical evidence. According to observations (Boylan-Kolchin 2013) the median Milky Way mass within 260kpc is $M_{totMW}(260kpc) = 1.6 \times 10^{12} M_{Sun}$ with a 90% confidence interval of $[1.0 - 2.4] \times 10^{12} M_{Sun}$. Hence, the amount of the effective gravitational charge of the quantum vacuum is nearly the same as the predicted amount of the hypothetical dark matter within dark matter paradigm and "phantom" dark matter within the MOND paradigm of modified gravity. The intriguing question is if the quantum vacuum enriched with virtual gravitational dipoles, can explain phenomena, usually attributed to competing paradigms of dark matter and modified gravity.

Imagine that the existence of dark matter is confirmed and that the proposed gravitational effects of the quantum vacuum do not exist; even in such a case the mystery will remain, why the radial density of dark matter in a galaxy agrees so well with Eq. (17) and can be calculated by formula $\sqrt{m_\pi M_b}/\lambda_\pi$.

In a complex system like a galaxy, exact analytical solutions are impossible and two competing paradigms (dark matter and modification of gravity) heavily depend on numerical methods and simulations. Of course the same is valid for this third emerging paradigm, according to which there is no dark matter and there is no modification of gravity, but quantum vacuum (as inherent part of the Standard Model of Particles and Fields) acts as a so far neglected ("forgotten") source of gravity. In order to give a fair chance to a new paradigm it must be treated with equal footing as the other two paradigms with intensive use of numerical methods and simulations.

### 6.1 An Intriguing Comparison with MOND

MOND is an ad-hoc theory (for a review see Famaey & McGaugh 2012), which is very successful for individual galaxies. Even the biggest proponents of dark matter paradigm admit the "*unreasonable effectiveness*" of MOND at galactic scale. Hence, if a theory significantly differs from MOND in description of a galaxy, it would be a serious sign that theory is wrong. Good agreement between a theory and MOND on galactic scale is a good sign for a new theory.

The starting point of MOND is an ad-hoc assumption that, for a point-like source of gravity, the ratio of the total and Newtonian acceleration ($g_{tot}/g_N$) is a function of the ratio ($a_0/g_N$) of a universal acceleration $a_0$ and the Newtonian acceleration, i.e.

$$\frac{g_{tot}}{g_N} = f\left(\frac{a_0}{g_N}\right) > 1. \tag{29}$$

In order to fit observations two limits are imposed on the function $f(a_0/g_N)$.

$$f\left(\frac{a_0}{g_N}\right) \to 1 \text{ when } \frac{a_0}{g_N} \to 0; f\left(\frac{a_0}{g_N}\right) \to \sqrt{\frac{a_0}{g_N}} \text{ when } \frac{a_0}{g_N} \to \infty. \tag{30}$$

Different interpolating functions $f(a_0/g_N)$ are used; the most popular ones are the simple, standard and RAR (the Radial Acceleration Relation) interpolating functions (Rodrigues et al. 2018):

$$f_{smp}\left(\frac{a_0}{g_N}\right) = \frac{1}{2}\left(1 + \sqrt{4\frac{a_0}{g_N} + 1}\right). \tag{31a}$$

$$f_{std}\left(\frac{a_0}{g_N}\right) = \frac{1}{\sqrt{2}}\sqrt{1 + \sqrt{4\left(\frac{a_0}{g_N}\right)^2 + 1}}. \tag{31b}$$

$$f_{rar}\left(\frac{a_0}{g_N}\right) = \frac{1}{1 - e^{-\sqrt{g_N/a_0}}}. \tag{31c}$$

There is a simple way to show that (at galactic scale) there is good agreement between MOND and our theory. In order to see it let us use equations (13) and (14) in order to get the total acceleration:

$$g_{tot} = g_N + g_{qv} = g_N + g_{qvmax} \tanh\left(\frac{R_{sat}}{r}\right) = g_N\left[1 + \frac{g_{qvmax}}{g_N}\tanh\left(\frac{R_{sat}}{r}\right)\right]. \tag{32}$$

From the point of view of MOND the term in square brackets is an interpolating function apparently very different from the best interpolating functions (3.5) used in MOND.

It is obvious that ratios $g_{qvmax}/g_N$ and $R_{sat}/r$ are respectively proportional to $a_0/g_N$ and $\sqrt{g_N/a_0}$; consequently the interpolating function can be written in the form which is easy for comparison with the above MOND interpolating functions.

$$f_{qv}\left(\frac{a_0}{g_N}\right) = 1 + \frac{g_{qv}}{g_N}\tanh\left(\frac{R_{sat}}{r}\right) = 1 + \alpha_1\frac{a_0}{g_N}\tanh\left(\alpha_2\sqrt{\frac{g_N}{a_0}}\right). \tag{33}$$

While we know roughly the dimensionless constants of proportionality $\alpha_1$ and $\alpha_2$, we have preferred to write Eq. (33) in more general form. For instance, with the value that we have adopted for $g_{qvmax} \approx 5 \times 10^{-11} \, m/s^2$, $\alpha_1 = g_{qvmax}/a_0 \approx 5/12$, but in general the numerical value is expected to be between 0.4 and 0.5.

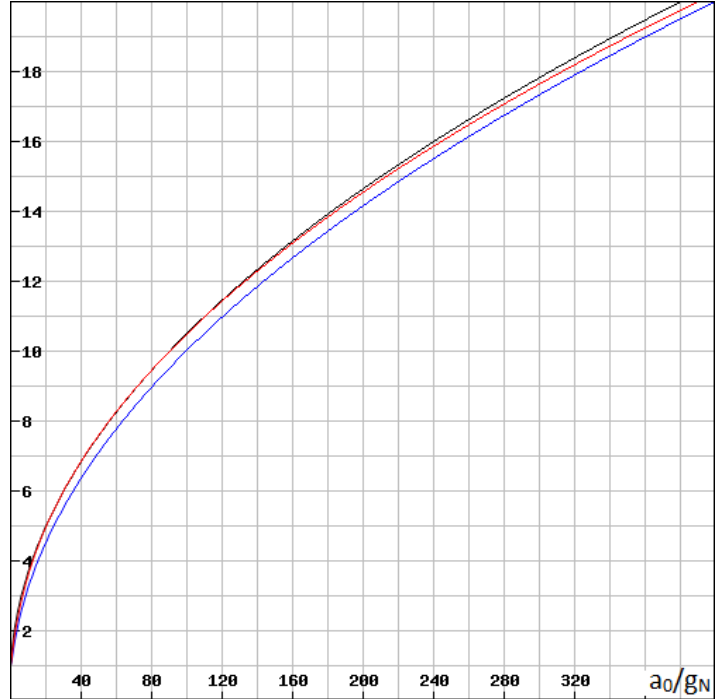

**Figure 6**: MOND's interpollating functions $f_{std}(a_0/g_N)$ in blue and $f_{rar}(a_0/g_0)$ in black, compared with the interpolating function $f_{qv}(a_0/g_0)$ "coming" from the quantum vacuum (with $\alpha_1 = 5/12$ and $\alpha_2 = 2.32$).

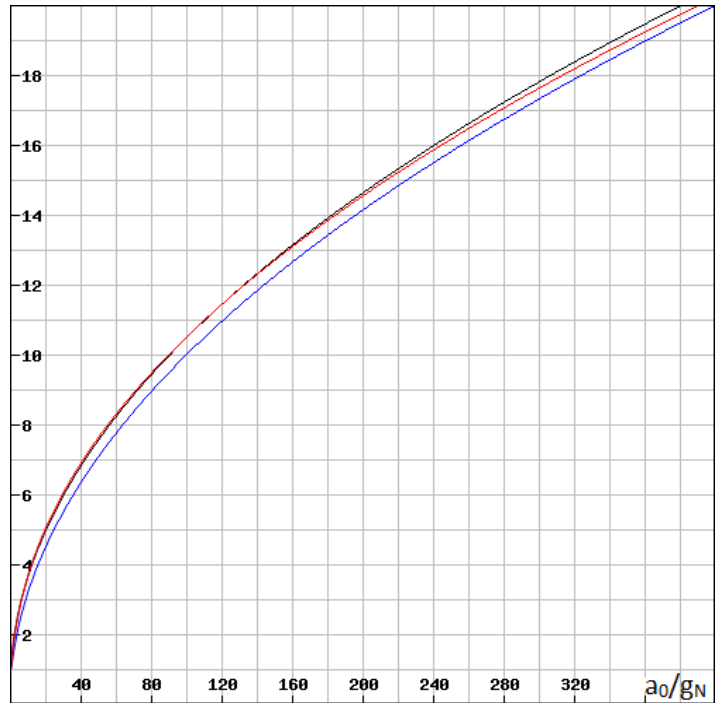

**Figure 7**: MOND's interpollating functions $f_{std}(a_0/g_N)$ in blue and $f_{rar}(a_0/g_0)$ in black, compared with the interpolating function $f_{qv}(a_0/g_0)$ "coming" from the quantum vacuum (with $\alpha_1 = 1/2$ and $\alpha_2 = 1.93$).

Figures 6 and 7, presents MOND's interpolating functions, the RAR function (Eq. (31c)) in red and the standard function (Eq. (31b)) in blue; the simple function (Eq, (31a)) is not shown only because in this small graph it would be indistinguishable from the RAR function. The green line *between* them shows our interpolating function Eq. (33) "coming" from the quantum vacuum. Apparently, from a numerical point of view our function is as good as MOND's functions, but from the fundamental point of view is superior.

# 7  Outlook and Discussion

My guess is that, in the way proposed in this paper or in a partially (or completely) different way, the quantum vacuum will be the cornerstone of the next scientific revolution.

If the existence of virtual gravitational dipoles is eventually confirmed, we will be forced to abandon the Standard ΛCDM Cosmology and to develop a new model of the Universe based on two fundamental principles:

(1) *The Standard Model matter* (i.e. matter made from quarks and leptons interacting through the exchange of gauge bosons) *is the only content of the Universe*.

(2) *Quantum vacuum fluctuations are virtual gravitational dipoles* (i.e. systems composed from one positive and one negative gravitational charge).

The first hypothesis excludes dark matter and dark energy from astrophysics and cosmology, while the second hypothesis postulates the quantum vacuum as a cosmological fluid free of the cosmological constant problem.

A huge majority of theoretical physicists (perhaps too huge to be right) is convinced that negative gravitational charges (and hence gravitational dipoles) cannot exist. However, if this scepticism of the majority is confirmed by the forthcoming empirical evidence, the results of this paper may still remain an encouraging and stimulating demonstration of how, in understanding the secrets of Nature, physical imagination and thinking are superior to purely mathematical thinking that has dominated physics for the last 40 years.

Our current understanding of the Universe is both, a fascinating intellectual achievement and the source of *the greatest crisis in the history of physics*. The first (and welcome) source of crisis are sophisticated astronomical observations that have revealed a series of phenomena that are a complete surprise and a complete mystery for contemporary physics. The second (and unwanted) source of crisis has been de facto suppression of alternative thinking, by dominating group-thinking.

In order to explain observations, besides the Standard Model matter, we have filled the Universe with hypothetical dark matter and dark energy, while in the primordial Universe we have assumed the existence of a mysterious inflation field (that causes a monstrous initial accelerated expansion of the Universe) and an enormous CP violation of unknown nature. And, after all these hypotheses, we still have not a plausible idea of what is the source of the cosmological constant problem.

Hence, we invoked a series of ad-hoc hypotheses and forced theories (the best example is supersymmetry) which, despite respectable mathematical beauty and value, are much more a result of mathematical than physical thinking. We do not know if all these hypotheses are correct or just well mimic something what we didn't understand. In any case, the current theoretical thinking is a departure from the traditional elegance, simplicity and beauty of theoretical physics.

The quantum vacuum is one of the most fundamental (if not the most fundamental) of all discoveries in the 20$^{th}$ century. It is unbelievable that, as a way out of crises, we have proposed so much of the unknown (from cosmic inflation, dark matter and dark energy to supersymmetry), without any serious attempt to use the quantum vacuum as a known and fundamental content of the Universe. Hopefully, even if our paper is wrong, it will motivate and encourage physicists, astrophysicists and cosmologists to think about the gravitational impact of the quantum vacuum.

Of course, many questions remain open and, because of limited space, many interesting topics were not discussed. Let us give just one intriguing example.

First, let us remember that before the emergence of structure (birth of the first stars and so on), the Universe was a very rarefied gas mainly composed of hydrogen and helium. Second, let us remember a crucial result revealed by the study of the Cosmic Microwave Background (CMB): the total real (or effective) mass (or as we prefer to say the gravitational charge of the Universe) was at the time of the birth of CMB, about 6.3 times larger than the baryonic gravitational charge. Within the ΛCDM Cosmology this additional gravitational charge is attributed to dark matter; consequently, if dark matter exists the ratio of the total amount of dark matter and baryonic matter in the Universe must be $M_{dmU}/M_{bU} \approx 5.3$. Strictly speaking the CMB can tell us the ratio only at that time but the ΛCDM assumption is that this ratio is the same today as well. By the way it means that the amount of dark matter in the Universe is a constant and hence dark matter as a cosmological fluid is a pressureless fluid.

Now let us turn to the picture that follows from the gravitational polarization of the quantum vacuum (See Figure 4 and discussion after Eq. (7)). The key point is that (at the time of the birth of the CMB) the mean distance between atoms of the cosmic gas is of the order of one millimetre, i.e. about 10 orders of magnitude (See Eq. (11)) larger than saturation radius of individual atoms. Hence, *there is enough space for each atom* (or nucleus if atoms are ionized) *to form a halo of the maximum size*; the total number of halos in the Universe is equal to the total number of atoms. Consequently, *at that time*, the total gravitational charge of each atom is a sum of its baryonic gravitational charge $\mathbf{M_b}$ and the effective gravitational charge of its halo (of the polarized quantum vacuum) of the maximum size, i.e. $\mathbf{M_{tot}(M_b) = M_b + M_{qvmax}(M_b)}$. Hence, if dark matter doesn't exist, and, if phenomena wrongly attributed to dark matter are caused by the gravitational polarization of the quantum vacuum, than, the empirical evidence $M_{dmU}/M_{bU} \approx \mathbf{5.3}$ (which is valid at the time of the birth of the CMB) must be reinterpreted as

$$\frac{\mathbf{M_{qvmax}(M_b)}}{\mathbf{M_b}} \approx \mathbf{5.3} \,. \tag{34}$$

We can alternatively say that, in the case of halos of the maximum size, each atom behaves as if its mass (gravitational charge) is multiplied by $\approx 6.3$. If this interpretation of the CMB data is correct, the quantum vacuum can replace dark matter in the process of structure formation in the Universe.

Let us underscore that ratio (34) that is valid for a point-like body *can be significantly larger* for structures (for instance a galaxy) composed of point-like bodies; hence the ratio increases with development of structures, but we will not discuss it here.

The scientific mainstream deserves *enormous* credit for detailed development of knowledge between two scientific revolutions, but the history of science teaches us that the mainstream is always surprised with scientific revolutions and in fact opposes them. In order to encourage the open-minded and imaginative thinking and critical attitude towards the prescribed truth let me end with an amusing law that is apparently valid in the time of scientific revolutions: *If you think differently from the mainstream it is not a proof that you are right, but if you think as the mainstream it is a proof that you are wrong.*

## Appendix A: Virtual gravitational dipoles and the universality of free fall

In a few years, experiments with antihydrogen will end a ninety-year-old mystery and reveal if antimatter (in the gravitational field of the Earth) falls just like ordinary matter or, antimatter falls upwards.

A huge majority of physicists believe that the outcome of these experiments is *known in advance*, i.e. that antimatter falls exactly in the same way as matter. This conviction is supported by apparently plausible arguments (for a review see Nieto & Goldman 1991, Chardin & Manfredi 2018) against the gravitational repulsion between matter and antimatter. However, there is also a fascinating argument (Villata 2011) that General Relativity and CPT symmetry (a cornerstone of the Standard Model of Particles and Fields) are compatible *only if* matter and antimatter mutually repel. Of course, only experiments can tell us who is right.

As already stated, the present paper is limited to the study of consequences of the working hypothesis that "*quantum vacuum fluctuations are virtual gravitational dipoles*", because, in our opinion, it is more important and productive than purely theoretical discussion whether repulsive gravity and virtual gravitational dipoles can exist or not. However, for completeness, in this Appendix, we present the main theoretical argument against the existence of virtual gravitational dipoles; an argument (see, instance, Cassidy 2018) based on a model-dependent theoretical interpretation of the experimental fact that nucleons (protons and neutrons) have complex structure dominated by virtual quark-antiquark pairs and gluons.

Historically the first (and naïf) structure of nucleons, with neglected quantum fluctuations, is presented on the left-hand side of Fig. A.1. Protons and neutrons are composed of 3 so called valence quarks of different strong (colour) charge which interact through the exchange of gluons presented with spirals; a proton is composed of one *d* and two *u* quarks, while a neutron contains one *u* and two *d* quarks. As experiments show such protons and neutrons do not exist in nature; the real structure of a proton (and similarly of a neutron), when quantum fluctuations are considered, is presented on the right-hand side of Fig. A.1.

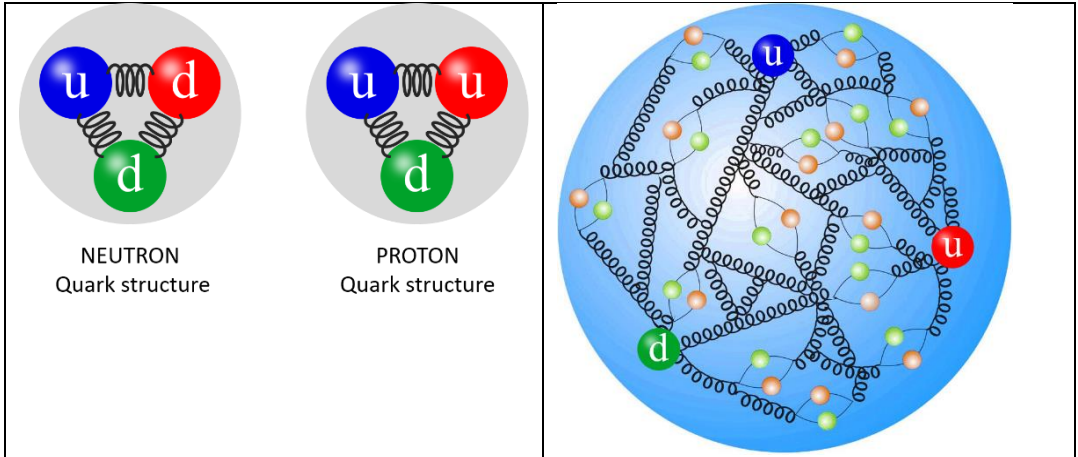

Figure A1: (a) Left-hand side presents structure of protons and neutrons with neglected quantum fluctuations. (b) Right-hand side shows inner structure of a proton revealed at HERA. Spirals represent gluons, while purple-green particles denote virtual quark-antiquark pairs. Note that all this is in addition to three valence quarks, two up and one down. (Source: DESY in Hamburg)

In brief, according to experiments, nucleons have a complex structure dominated by virtual content; the estimated mass of 3 valence quarks (i.e. non-virtual content) inside a nucleon is about two orders of magnitude smaller than the measured mass of the nucleon. A difficult task within quantum chromodynamics (QCD) is the transition from this qualitative picture to the quantitative understanding of how the observed masses of nucleons emerge from the confinement and dynamics of the aforementioned content (valence quarks, quark-antiquark pairs and gluons). Currently, the most powerful numerical method known as lattice QCD predicts relatively accurately (Yang et al. 2018, Walker-Loud 2018)) the distinct contributions to the total mass of nucleons.

Now, it is easy to understand the most important theoretical argument against the existence of virtual gravitational dipoles. If virtual gravitational dipoles exist inside nucleons, they contribute differently to the mass and gravitational charge of nucleons; hence, the well-established universality of free fall must be violated. Of course, this theoretical argument must be taken seriously, but as it will become clear below, it is model-dependent, and its validity is uncertain.

At this point it is crucial to remember that the Standard Model of Particles and Fields doesn't incorporate gravitational interactions; in cases when we are interested in gravitation, or gravitation cannot be neglected, we simply combine the Standard Model with our current theory of gravitation. We already know that sometimes, such a combination of the Standard Model and our theory of gravitation, leads to extremely wrong predictions; for instance, if the mass-energy density of the quantum vacuum in the Standard Model is interpreted as the gravitational charge (as it must be according to our current understanding of gravitation) it leads to the cosmological constant problem i.e. to the worst prediction in the history of physics. We can say that the appearance of the cosmological constant problem is model-dependent; as already noted the problem doesn't exist within the model of virtual gravitational dipoles. Hence, we must be very careful in the application of our "gravitational reasoning" to experimental evidence coming from Standard Model research. In particular, the empirical evidence for complex structure of nucleons tells us nothing about their gravitational charge; the gravitational charge of a nucleon cannot be calculated without the additional theoretical assumptions about the unknown gravitational charges of its constituents, i.e. calculation is *model dependent*. Just as the simplest illustration of different theoretical possibilities and our limited understanding of the gravitational properties of quarks and gluons, let us note that we do not know if the Weak Equivalence Principle (i.e. the equivalence between inertial mass and gravitational charge) is valid for quarks and gluons. For instance, while it seems quite plausible, we do not know if, for the $d$ and $u$ quark, there is equality between their mass ratio and their gravitational charge ratio. With such an incomplete knowledge, all theoretical arguments are no more than a model-dependant gravitational interpretation of non-gravitational empirical evidence.

I would like to thank an excellent anonymous Referee whose questions and comments motivated the writing of this Appendix, as well as the Appendix C.

## Appendix B: Reflections on Black-White Holes

This Appendix is just an illustration of diversity of consequences of an eventual gravitational repulsion between matter and antimatter.

The Universe is full of black holes. Just in our galaxy, in addition to the central supermassive black hole, there are perhaps about $10^7$ stellar mass black holes with an average mass below 10 Solar masses. While no one thinks about it, within a few years, experiments at CERN might "convert" all black holes into black-*white* holes!

In fact, as already noticed, after more than one decade of complex preparation, experiments at CERN (ALPHA, AEGIS and GBAR) will measure the *gravitational acceleration of antihydrogen* in the gravitational field of the Earth. You may be surprised that nine decades after the discovery of antimatter, we cannot answer the simplest question: In which direction would an anti-apple fall in the gravitational field of the Earth, down or up? We know that an apple falls down, but, no one knows if an anti-apple would also fall down or would fall upwards. While a huge majority of theoretical physicists (perhaps too huge to be right) believe that the result of experiments is known in advance, i.e. that antimatter falls in the same way as matter, it may be a good idea to wait and see.

Let us assume that experiments confirmed *the gravitational repulsion between matter and antimatter*. So far, no one has noticed that it would be a proof of the existence of *black-white* holes; *black-white* holes are inevitable if antihydrogen falls upwards in the gravitational field of the Earth. What is a *black-white* hole? Well, if matter and antimatter gravitationally repel each other, black holes must be renamed *black-white* holes; a black hole made from matter is a black hole for matter but a *white* hole for antimatter. Matter *cannot escape* if it is inside the horizon, while antimatter because of repulsion *cannot remain* inside the horizon. Similarly (according to CPT symmetry) a black hole made of antimatter is a black hole for antimatter, but a white hole for matter.

Let us consider the example of a Schwarzschild black hole made of matter. It is obvious that the metric "seen" by a test particle and the metric "seen" by a test antiparticle are respectively:

$$ds^2 = c^2 \left(1 - \frac{2GM}{c^2 r}\right) dt^2 - \left(1 - \frac{2GM}{c^2 r}\right)^{-1} dr^2 - r^2(d\theta^2 + \sin^2\theta \, d\phi). \tag{B.1}$$

$$ds^2 = c^2 \left(1 + \frac{2GM}{c^2 r}\right) dt^2 - \left(1 + \frac{2GM}{c^2 r}\right)^{-1} dr^2 - r^2(d\theta^2 + \sin^2\theta \, d\phi). \tag{B.2}$$

The key point is that according to metric (B.1) there is a horizon for matter defined by the Schwarzschild radius $R_S = 2GM/c^2$, while according to metric (B.2) there is no horizon for antimatter.

### B.1 Black-white holes – a source of antimatter in cosmic rays

If particles and antiparticles have *gravitational charge of the opposite sign*, in our Universe dominated by matter, *black-white* holes must be a source of antiprotons and positrons in cosmic rays.

Let us consider matter falling into a *black-white* hole. The total energy of a falling particle is its rest energy $mc^2$ plus energy $GMm/r$ gained by gravitational acceleration (note the use of metric (1)). The energy gained by free-fall becomes equal to the rest energy at $r = R_S/2$; hence, only in the inner part of the matter horizon, the total energy of the particle becomes larger than $2mc^2$ (a threshold for creation of particle-antiparticle pairs in collisions). As a result of collisions of infalling particles (analogous to collisions in our accelerators), different kinds of antiparticles can be created inside the matter horizon and, long-living antiparticles would be violently ejected outside the horizon (in fact positrons and antiprotons and eventually antineutrinos if gravitational repulsion is valid for them as well). Of course, ejection rate should be greater if the quantity of the infalling material is greater; a black-white hole behaves as an irregular source of antimatter.

According to our best knowledge and experience in production of antiprotons, only a miniscule fraction of falling matter will be converted to antimatter and ejected back to space. *It would be highly important to perform computer simulations to get an insight into black-white holes as one possible source of antimatter in cosmic rays*.

An intriguing question is if two different signatures of these black-white holes in our galaxy have already been seen! The first signature may be an unexplained excess of high-energy positrons and antiprotons in cosmic rays (Accardo et al. 2014) revealed by the measurements with the Alpha Magnetic Spectrometer on the International Space Station. The second signature may be recent detection, at the IceCube neutrino telescope at the South pole, of very high-energy (anti)neutrinos coming from the galactic centre (Bai *et al*. 2014); apparently the Milky Way's supermassive black hole acts as a mysterious "factory" of high-energy (anti)neutrinos.

## B.2 Black-white hole radiation

There is a second, more subtle mechanism for creation of particle-antiparticle pairs deep inside the horizon. Let us remember again that the quantum vacuum is an inherent part of the Standard Model of Particles and Fields (e.g. Aitchison 2009) and that under certain conditions virtual particle-antiparticle pairs from the quantum vacuum can be converted into real particles; we can create something from apparently nothing. For instance, an electron and positron in a virtual pair can be converted to real ones in a sufficiently strong electric field (Schwinger 1951) accelerating them in opposite directions. The same (i.e. creation of particle-antiparticle pairs from the quantum vacuum) can be done by a gravitational field if particles and antiparticles have gravitational charge of the opposite sign; the only difference is that the needed opposite acceleration is caused by a gravitational field. The particle-antiparticle creation rate per unit volume and time is given by relation (B.3); creation of antiparticles is significant only if the gravitational acceleration $g$ is larger than a critical value $g_{cr}$.

$$\frac{dN_{m\bar{m}}}{dtdV} \approx \frac{c}{\lambda^4}\left(\frac{g}{g_{cr}}\right)^2 \sum_{n=1}^{\infty}\frac{1}{n^2}exp\left(-n\frac{g_{cr}}{g}\right), g_{cr} \equiv \pi\frac{c^2}{\lambda_m}, \quad \lambda_m \equiv \frac{\hbar}{mc}. \tag{B.3}$$

Let us note that Eq. (B.3) is the gravitational version (Hajdukovic 2014) of the well-known Schwinger mechanism (Schwinger 1951).

Hence, black-white holes might radiate because of particle-antiparticle creation from the quantum vacuum.

Can Hawking radiation coexist with quantum vacuum radiation? No. Hawking radiation depends on the heretofore accepted model of the gravitational properties of the quantum vacuum. Hawking calculations correspond to the case of gravitational monopoles and cannot be valid if the quantum vacuum is composed of gravitational dipoles.

# Appendix C: The Schwarzschild metric with the gravitational polarization of the quantum vacuum

So far, the gravitational field of a point-like body immersed in the quantum vacuum was studied in the framework of the Newtonian theory of gravity.

The Schwarzschild metric (Eq. (2)) is the general-relativistic description of the gravitational impact of a point-like body immersed in the gravitationally featureless vacuum. Newtonian gravity (with the quantum vacuum neglected) can be considered as weak-field limit of Schwarzschild gravity. In an analogous way, Newtonian gravity, with the included gravitational impact of the quantum vacuum is the week-field limit of a more general metric:

$$ds^2 = c^2\left(1 - \frac{R_S}{r} - \frac{8\pi G}{c^2}rP_g(M_b, r)\right)dt^2 - \left(1 - \frac{R_S}{r} - \frac{8\pi G}{c^2}rP_g(M_b, r)\right)^{-1}dr^2 - r^2d\Omega^2 \tag{C.1}$$

Let us remember that $0 \leq P_g(M_b, r) \leq P_{gmax}$, and, that in the region of saturation we can use the equality $P_g(M_b, r) = P_{gmax}$. Additionally, $R_S \ll R_{sat}$, i.e. the Schwarzschild radius of the point-like body is many orders of magnitude smaller than the size of the corresponding region of saturation (See Eq. (11)); hence, with a very high accuracy, the region outside the region of saturation can be considered as the weak-field limit.

Let us underscore that *deep* inside the region of saturation

$$\frac{R_S}{r} \gg \frac{8\pi G}{c^2} r P_g(M_b, r)$$

For instance, in the case of the Sun, at the distance of Mercury, the left-hand side is nine orders of magnitude larger than the right-hand side; consequently, there is only a tiny (and with our current precision non-measurable) contribution of the quantum vacuum to the already known general-relativistic description of the orbit of Mercury.

## REFERENCES

Accardo L. et al. (AMS Collaboration). High Statistics Measurement of the Positron Fraction in Primary Cosmic Rays of 0.5–500 GeV with the Alpha Magnetic Spectrometer on the International Space Station. Phys. Rev. Lett. 113, 121101 (2014)

Aitchison I.J.R., Nothing's plenty - the vacuum in modern quantum field theory, 2009, Contemp. Physics, 50, 261–319.

ALPHA Collaboration., Confinement of antihydrogen for 1,000 seconds, 2011, Nature Physics, 7, 558–564.

ALPHA Collaboration., Observation of the 1S–2S transition in trapped antihydrogen, 2017, *Nature*, 541, 506–510.

Bai Y. et al. Neutrino lighthouse at Sagittarius A*. Phys. Rev. D 90, 063012 (2014)

Bertsche W.A., Prospects for comparison of matter and antimatter gravitation with ALPHA-g, 2018, Trans. R. Soc. A, 376, 20170265.

Boylan-Kolchin M., The Space Motion of Leo I: The Mass of the Milky Way's Dark Matter Halo, 2013, The Astrophysical Journal, 768, 140.

Brusa R.S. et al., The AEGIS experiment at CERN, 2017, J. Phys.: Conf. Ser., 791, 012014.

Cassidy D.B., Hogan S.D., Atom control and gravity measurements, 2014, Int. Journal of Modern Physics: Conf. Ser., 30, 1460259.

D.B. Cassidy, Experimental progress in positronium laser physics, Eur. Phys. J. D 72, 52 (2018)

G. Chardin & G. Manfredi, Gravity, antimatter and the Dirac-Milne universe. Hyperfine Interactions 239, 45 (2018).

Dirac P.A.M., The Cosmological Constants, 1937 Nature, 139, 323.

Dirac P.A.M., A new basis for cosmology, 1938, Proc. R. Soc. A, 165, 199.

Famaey B. & McGaugh S.S., Modified Newtonian Dynamics (MOND), 2012, Living Rev. Relastivity 15, 10

Fienga A. et al., Gravity tests with INPOP planetary ephemerides, 2009, Proceedings IAU Symposium, 261, 59-169,

Gai M., Vecchiato A., Astrometric detection feasibility of gravitational effects of quantum vacuum, 2014, Arxiv, http://arxiv.org/abs/1406.3611v2 .

Hajdukovic D.S., On the relation between mass of a pion, fundamental physical constants and cosmological parameters, 2010, EPL, 89, 49001

Hajdukovic D.S., Is dark matter an illusion created by the gravitational polarization of the quantum vacuum, 2011, Astrophysics and Space Science, 334, 215-218.

Hajdukovic D.S., Virtual gravitational dipoles: The key for the understanding of the Universe, 2014, Physics of the Dark Universe, 3, 34-40.

L3 Collaboration., Measurement of the running of the fine-structure constant, 2000, Phys. Lett. B, 476, 40–48.

Milgrom M., MOND – a pedagogical review, 2011, Preprint at arXiv:astro-ph/0112069v1

Murray C.D., Dermott S.F., *Solar System Dynamics*, 1999, Cambridge University Press.

Nieto M.M. & Goldman T., The arguments against "antigravity" and the gravitational acceleration of antimatter. Physics Reports 205, 221-281 (1991).

Peebles P.J.E., Growth of the nonbaryonic dark matter theory, 2017, Nature Astronomy, 1, 0057.

Perez P. et al., The GBAR antimatter gravity experiment, 2015, Hyperfine Interactions, 233, 21-27.

Phillips T.J., The Muonium Antimatter Gravity Experiment, 2018, EPJ Web of Conferences, 181**,** 01017.

Pitjev, N.V.; Pitjeva, E.V. Constraints on Dark Matter in the Solar System, 2013, Astronomy Letters, 39, 141-149.

Rodrigues D.C. et al., Absence of a fundamental acceleration scale in galaxies, 2018, Nature Astronomy, 2, 668-672

Schwinger, J.S., On gauge invariance and vacuum polarization, 1951, Phys. Rev., 82, 664–679.

Villata M., CPT symmetry and antimatter gravity in General Relativity. EPL 94, 20001 (2011).

Walker-Loud, A., Dissecting the Mass of the Proton, 2018, Physics, 11, 118.

Weinberg S., *Gravitation and Cosmology*, 1972, John Wiley and Sons, pp 620.

Weinberg, S., The cosmological constant problem, 1989, Rev. Mod. Phys., 61, 1–23.

Wilson C.M. et al., Observation of the dynamical Casimir effect in a superconducting circuit, 2011, Nature, 479, 376–379.

Yang Y.B. et al., Proton mass decomposition from the QCD energy momentum tensor, 2018, Phys. Rev. Lett. 121, 212001.